\documentclass[bibtex,twocolumn,prl,showpacs,floatfix,preprintnumbers,amsmath,amssymb,superscriptaddress,letter]{revtex4}

\usepackage{graphicx}
\usepackage{dcolumn}
\usepackage{color}
\usepackage{xspace}
\usepackage[tight]{units}

\usepackage{fancyhdr}
\newcommand{\numu}{\ensuremath{\nu_{\mu}}\xspace}
\newcommand{\numubar}{\ensuremath{\overline{\nu}_{\mu}}\xspace}

\newcommand{\st}{\ensuremath{\sin^{2}2{\theta}}}
\newcommand{\eps}{\ensuremath{\varepsilon_{\mu\tau}}}

\newcommand{\dmsq}{\ensuremath{|\Delta m^{2}|}}

\makeatletter

\newcommand{\Rmnum}[1]{\expandafter\@slowromancap\romannumeral #1@}
\makeatother

\newcommand{\epsresult}{\ensuremath{-0.20 < \varepsilon_{\mu\tau} < 0.07\;\text{(90\% C.L.)}}}

\begin{document}

\vspace*{.6cm} 
\title{A search for flavor-changing non-standard neutrino interactions by MINOS}




\newcommand{\Berkeley}{Lawrence Berkeley National Laboratory, Berkeley, California, 94720 USA}
\newcommand{\Cambridge}{Cavendish Laboratory, University of Cambridge, Madingley Road, Cambridge CB3 0HE, United Kingdom}
\newcommand{\Cincinnati}{Department of Physics, University of Cincinnati, Cincinnati, Ohio 45221, USA}
\newcommand{\FNAL}{Fermi National Accelerator Laboratory, Batavia, Illinois 60510, USA}
\newcommand{\RAL}{Rutherford Appleton Laboratory, Science and Technologies Facilities Council, OX11 0QX, United Kingdom}
\newcommand{\UCL}{Department of Physics and Astronomy, University College London, Gower Street, London WC1E 6BT, United Kingdom}
\newcommand{\Caltech}{Lauritsen Laboratory, California Institute of Technology, Pasadena, California 91125, USA}
\newcommand{\Alabama}{Department of Physics and Astronomy, University of Alabama, Tuscaloosa, Alabama 35487, USA}
\newcommand{\ANL}{Argonne National Laboratory, Argonne, Illinois 60439, USA}
\newcommand{\Athens}{Department of Physics, University of Athens, GR-15771 Athens, Greece}
\newcommand{\NTUAthens}{Department of Physics, National Tech. University of Athens, GR-15780 Athens, Greece}
\newcommand{\Benedictine}{Physics Department, Benedictine University, Lisle, Illinois 60532, USA}
\newcommand{\BNL}{Brookhaven National Laboratory, Upton, New York 11973, USA}
\newcommand{\CdF}{APC -- Universit\'{e} Paris 7 Denis Diderot, 10, rue Alice Domon et L\'{e}onie Duquet, F-75205 Paris Cedex 13, France}
\newcommand{\Cleveland}{Cleveland Clinic, Cleveland, Ohio 44195, USA}
\newcommand{\Delhi}{Department of Physics \& Astrophysics, University of Delhi, Delhi 110007, India}
\newcommand{\GEHealth}{GE Healthcare, Florence South Carolina 29501, USA}
\newcommand{\Harvard}{Department of Physics, Harvard University, Cambridge, Massachusetts 02138, USA}
\newcommand{\HolyCross}{Holy Cross College, Notre Dame, Indiana 46556, USA}
\newcommand{\Houston}{Department of Physics, University of Houston, Houston, Texas 77204, USA}
\newcommand{\IIT}{Department of Physics, Illinois Institute of Technology, Chicago, Illinois 60616, USA}
\newcommand{\Iowa}{Department of Physics and Astronomy, Iowa State University, Ames, Iowa 50011 USA}
\newcommand{\Indiana}{Indiana University, Bloomington, Indiana 47405, USA}
\newcommand{\ITEP}{High Energy Experimental Physics Department, ITEP, B. Cheremushkinskaya, 25, 117218 Moscow, Russia}
\newcommand{\JMU}{Physics Department, James Madison University, Harrisonburg, Virginia 22807, USA}
\newcommand{\LASL}{Nuclear Nonproliferation Division, Threat Reduction Directorate, Los Alamos National Laboratory, Los Alamos, New Mexico 87545, USA}
\newcommand{\Lebedev}{Nuclear Physics Department, Lebedev Physical Institute, Leninsky Prospect 53, 119991 Moscow, Russia}
\newcommand{\LLL}{Lawrence Livermore National Laboratory, Livermore, California 94550, USA}
\newcommand{\LosAlamos}{Los Alamos National Laboratory, Los Alamos, New Mexico 87545, USA}
\newcommand{\Manchester}{School of Physics and Astronomy, University of Manchester, Oxford Road, Manchester M13 9PL, United Kingdom}
\newcommand{\MIT}{Lincoln Laboratory, Massachusetts Institute of Technology, Lexington, Massachusetts 02420, USA}
\newcommand{\Minnesota}{University of Minnesota, Minneapolis, Minnesota 55455, USA}
\newcommand{\Crookston}{Math, Science and Technology Department, University of Minnesota -- Crookston, Crookston, Minnesota 56716, USA}
\newcommand{\Duluth}{Department of Physics, University of Minnesota Duluth, Duluth, Minnesota 55812, USA}
\newcommand{\Ohio}{Center for Cosmology and Astro Particle Physics, Ohio State University, Columbus, Ohio 43210 USA}
\newcommand{\Otterbein}{Otterbein College, Westerville, Ohio 43081, USA}
\newcommand{\Oxford}{Subdepartment of Particle Physics, University of Oxford, Oxford OX1 3RH, United Kingdom}
\newcommand{\PennState}{Department of Physics, Pennsylvania State University, State College, Pennsylvania 16802, USA}
\newcommand{\PennU}{Department of Physics and Astronomy, University of Pennsylvania, Philadelphia, Pennsylvania 19104, USA}
\newcommand{\Pittsburgh}{Department of Physics and Astronomy, University of Pittsburgh, Pittsburgh, Pennsylvania 15260, USA}
\newcommand{\IHEP}{Institute for High Energy Physics, Protvino, Moscow Region RU-140284, Russia}
\newcommand{\Rochester}{Department of Physics and Astronomy, University of Rochester, New York 14627 USA}
\newcommand{\RoyalH}{Physics Department, Royal Holloway, University of London, Egham, Surrey, TW20 0EX, United Kingdom}
\newcommand{\Carolina}{Department of Physics and Astronomy, University of South Carolina, Columbia, South Carolina 29208, USA}
\newcommand{\SLAC}{Stanford Linear Accelerator Center, Stanford, California 94309, USA}
\newcommand{\Stanford}{Department of Physics, Stanford University, Stanford, California 94305, USA}
\newcommand{\StJohnFisher}{Physics Department, St. John Fisher College, Rochester, New York 14618 USA}
\newcommand{\Sussex}{Department of Physics and Astronomy, University of Sussex, Falmer, Brighton BN1 9QH, United Kingdom}
\newcommand{\TexasAM}{Physics Department, Texas A\&M University, College Station, Texas 77843, USA}
\newcommand{\Texas}{Department of Physics, University of Texas at Austin, 1 University Station C1600, Austin, Texas 78712, USA}
\newcommand{\TechX}{Tech-X Corporation, Boulder, Colorado 80303, USA}
\newcommand{\Tufts}{Physics Department, Tufts University, Medford, Massachusetts 02155, USA}
\newcommand{\UNICAMP}{Universidade Estadual de Campinas, IFGW-UNICAMP, CP 6165, 13083-970, Campinas, SP, Brazil}
\newcommand{\UFG}{Instituto de F\'{i}sica, Universidade Federal de Goi\'{a}s, CP 131, 74001-970, Goi\^{a}nia, GO, Brazil}
\newcommand{\USP}{Instituto de F\'{i}sica, Universidade de S\~{a}o Paulo,  CP 66318, 05315-970, S\~{a}o Paulo, SP, Brazil}
\newcommand{\Warsaw}{Department of Physics, University of Warsaw, Ho\.{z}a 69, PL-00-681 Warsaw, Poland}
\newcommand{\Washington}{Physics Department, Western Washington University, Bellingham, Washington 98225, USA}
\newcommand{\WandM}{Department of Physics, College of William \& Mary, Williamsburg, Virginia 23187, USA}
\newcommand{\Wisconsin}{Physics Department, University of Wisconsin, Madison, Wisconsin 53706, USA}
\newcommand{\deceased}{Deceased.}

\affiliation{\ANL}
\affiliation{\Athens}
\affiliation{\BNL}
\affiliation{\Caltech}
\affiliation{\Cambridge}
\affiliation{\UNICAMP}
\affiliation{\Cincinnati}
\affiliation{\FNAL}
\affiliation{\UFG}
\affiliation{\Harvard}
\affiliation{\HolyCross}
\affiliation{\Houston}
\affiliation{\IIT}
\affiliation{\Indiana}
\affiliation{\Iowa}
\affiliation{\UCL}
\affiliation{\Manchester}
\affiliation{\Minnesota}
\affiliation{\Duluth}
\affiliation{\Otterbein}
\affiliation{\Oxford}
\affiliation{\Pittsburgh}
\affiliation{\RAL}
\affiliation{\USP}
\affiliation{\Carolina}
\affiliation{\Stanford}
\affiliation{\Sussex}
\affiliation{\TexasAM}
\affiliation{\Texas}
\affiliation{\Tufts}
\affiliation{\Warsaw}
\affiliation{\WandM}

\author{P.~Adamson}
\affiliation{\FNAL}











\author{G.~Barr}
\affiliation{\Oxford}









\author{M.~Bishai}
\affiliation{\BNL}

\author{A.~Blake}
\affiliation{\Cambridge}


\author{G.~J.~Bock}
\affiliation{\FNAL}


\author{D.~Bogert}
\affiliation{\FNAL}




\author{S.~V.~Cao}
\affiliation{\Texas}



\author{D.~Cherdack}
\affiliation{\Tufts}

\author{S.~Childress}
\affiliation{\FNAL}


\author{J.~A.~B.~Coelho}
\affiliation{\Tufts}
\affiliation{\UNICAMP}



% ...

\collaboration{The MINOS Collaboration}
\noaffiliation




\begin{abstract}
We report new constraints on flavor-changing non-standard neutrino interactions (NSI) using data from the MINOS experiment. We analyzed a combined set of beam neutrino and antineutrino data, and found no evidence for deviations from standard neutrino mixing.  The observed energy spectra constrain the NSI parameter to the range \epsresult.

\end{abstract} 

\maketitle
\thispagestyle{fancy}

It is well established from solar, atmospheric, reactor and accelerator experiments~\cite{ref:sk,ref:sno,ref:minos2006,ref:kamland,ref:borexino,ref:k2k,ref:db_reno} that neutrinos undergo flavor change as they propagate. This phenomenon can be explained by the quantum  mechanical mixing of neutrino flavor and mass eigenstates. The mixing can be parametrized by  three angles, $\theta_{12},\,\theta_{13},\,\theta_{23}$, and a CP-violating phase, $\delta$~\cite{ref:pmns}. The standard neutrino oscillation mechanism requires that at  least two of the three active neutrinos are massive. While the phenomenon of oscillations can occur in vacuum,  the presence of matter along the neutrino path allows for alternative flavor changing mechanisms such as the Mikheyev-Smirnov-Wolfenstein~(MSW) matter effect~\cite{ref:msw_original}. This effect alters the survival probability of electron neutrinos propagating through matter since electron neutrinos can have additional interactions with electrons in the surrounding medium, but does not affect the survival probability of muon or tau neutrinos. 

Non-standard interactions (NSI)~\cite{ref:nsi} could occur between muon or tau neutrinos and matter, and could alter the flavor content of a neutrino beam as it propagates through the Earth's crust in a manner similar to standard matter effects. Searches for NSI have already been performed with atmospheric neutrinos~\cite{ref:superK}. However, non-standard matter effects are, in general, different for neutrinos and antineutrinos. Accelerator-based oscillation experiments offer a powerful tool to search for NSI with their ability to produce well-understood beams of neutrinos and antineutrinos separately. Furthermore, MSW and NSI effects depend on the neutrino's path length in matter, and beam neutrinos travel a well defined distance to the detector. Recent papers have discussed the compatibility of NSI with MINOS data~\cite{ref:Mann_PRD,Kopp:2010qt}. This Letter describes the first direct search for NSI which simultaneously fits the separate energy spectra of neutrinos and antineutrinos in a long-baseline experiment. 

Short-baseline neutrino experiments have explored and constrained non-standard interactions~\cite{ref:nsi_sbl}. In such experiments both charged-current (CC) and neutral-current (NC) NSI can be studied. While long-baseline experiments can also constrain interactions with matter directly in either near or far detectors, their improved sensitivity arises from  using the Earth's matter along the long neutrino path as the interaction medium. Charged-current NSI produce a final state charged-lepton which is absorbed by the Earth's matter and not observed.  However, NC NSI that occur along the neutrino path will produce a final state neutrino of a different flavor, altering the flavor content of the beam~\cite{Kopp:2010qt}. We consider only NC NSI in this study. 

The NSI Hamiltonian can be included as a perturbation to standard oscillations. This Hamiltonian is proportional to the matter potential $V=\sqrt{2}G_{F}N_{e}$, with $G_{F}$, the Fermi coupling constant, and $N_{e}$, the electron density in matter, analogous to the MSW matter effect~\cite{ref:msw_original}. In general, the Hamiltonian has both flavor-conserving and flavor-changing components. In flavor-conserving NSI, the NC scattering between the neutrino and matter does not alter the neutrino flavor.  Flavor-changing NSI, on the other hand, do not conserve lepton flavor number; the final state neutrino is in a different flavor eigenstate from the initial neutrino. In a disappearance experiment, flavor-changing NSI have a greater effect on the flavor transition probability than flavor-conserving NSI due to interference between amplitudes of standard and non-standard matter interactions~\cite{ref:nsi_interference}. We consider here only flavor-changing NSI and set to zero flavor-conserving amplitudes to which MINOS has no sensitivity.

The resulting NSI Hamiltonian in the two-flavor approximation is
\begin{equation}
H_{\text{NSI}}=V
\left(\begin{array}{cc}
0 & \varepsilon_{\mu\tau}\\
\varepsilon^{*}_{\mu\tau} & 0
\end{array}\right),
\end{equation}

\noindent where the coefficient $\varepsilon_{\mu\tau}$ gives the strength of the NSI effect on transitions between $\mu$ and $\tau$ flavors. We only consider the real part of $\varepsilon_{\mu\tau}$, which is sensitive to differences between \numu\ and \numubar\ survival. We define the vacuum oscillation length for neutrinos of energy $E$ as $L_{0}\equiv\left(\frac{4E}{\Delta m^{2}}\right)$. The difference between the squares of the second and third neutrino masses, $\Delta m^{2} = \Delta m^{2}_{32} \equiv m^{2}_{3} - m^{2}_{2}$, is the same parameter that governs standard neutrino oscillations~\cite{ref:minosCC2010,comment:dmbar}. The NSI matter oscillation length is defined as
\begin{equation}
    L_{m} \equiv \frac{L_{0}}{\left[ 1\pm2\sin(2\theta)L_{0}\varepsilon_{\mu\tau} \lvert V \rvert + (L_{0}\varepsilon_{\mu\tau} \lvert V \rvert)^{2} \right]^{\frac{1}{2}}}.
    \label{eq:mat_length}  
  \end{equation}
\noindent The survival probability can then be written as
\begin{equation}
  P(\nu_{\mu}\rightarrow\nu_{\mu}) = 1- \left[ 1- \cos^{2}(2\theta)\frac{L_{m}^{2}}{L_{0}^{2}}\right]\sin^{2}\left(\frac{L}{L_{m}} \right),
  \label{eq:nsi_prob}  
\end{equation}
\noindent with mixing angle $\theta = \theta_{23} $, and neutrino path length $L$. 
Standard oscillation parameters $\theta$ and $\Delta m^{2}$ are taken to be the same for neutrinos and antineutrinos. The $\pm$ signs in Eq.~(\ref{eq:mat_length}) arise from the matter potential, $V$, which is positive for neutrinos and negative for antineutrinos. The parameter $\varepsilon_{\mu\tau}$ is real-valued and carries its own sign. A positive value of $\varepsilon_{\mu\tau}$ implies that the neutrino disappearance probability is greater than the antineutrino disappearance probability, and vice versa.

The MINOS experiment measures disappearance of muon neutrinos and antineutrinos in the  NuMI beam~\cite{ref:NuMI} using two detectors.  The event energy spectrum of the low energy NuMI beam, used in this measurement, peaks at approximately 3~GeV~\cite{ref:minosnim}. Its focusing components can be tuned to produce a beam with an event composition of 91.7\%~\numu, 7\%~\numubar, and 1.3\%~($\nu_{e}+\overline{\nu}_{e}$) in neutrino mode, or of 58\%~\numu, 40\%~\numubar, and 2\%~($\nu_{e}+\overline{\nu}_{e}$) in antineutrino mode~\cite{ref:rhc}. The Near Detector (ND), with a fiducial mass of 23.7~tons, measures the neutrino and antineutrino energy spectra $1.04$~km downstream of the production target. The Far Detector (FD) is located in the Soudan Underground Laboratory and has a 4.2~kiloton fiducial mass. It measures the energy spectra 735~km downstream of the production target. Both detectors are magnetized steel-scintillator tracking-sampling calorimeters designed to measure the energy and the sign of the charge of muons produced by \numu\ and \numubar\ interactions. In each detector,  muon neutrino and antineutrino CC interactions are separated event-by-event using 
the sign of muon track curvature.

The results presented here are based on an exposure of $7.09\times10^{20}$ protons on target 
(POT) in neutrino mode, combined with a $2.95\times10^{20}$ POT exposure in antineutrino mode. MINOS previously reported a two-flavor oscillation analysis on this neutrino dataset~\cite{ref:minosCC2010,comment:dmbar} and the first direct measurement of antineutrino oscillation parameters from the antineutrino sample~\cite{ref:rhc}. Due to the opposite sign of the matter potential in Eq.~(\ref{eq:nsi_prob}) for neutrinos and antineutrinos, NSI, if present, will alter the survival probability of neutrinos and antineutrinos in opposite directions. The magnitude of $\varepsilon_{\mu\tau}$ is proportional to the difference in probability between neutrinos and antineutrinos, and the sign of $\varepsilon_{\mu\tau}$ is determined by the sign of the probability difference.

We select \numu\ and \numubar\ CC events inside the fiducial volume by identifying interaction vertices with a muon track and possible hadronic activity. The neutrino energy is reconstructed by summing the muon track and hadronic shower  energies. Muon energy is measured using range for muons that stop in the detector and curvature for muons that exit. The hadronic energy is determined using a $k$-Nearest Neighbor ($k$NN) technique~\cite{ref:ChrisThesis}. We require the muon charge be  negative for \numu\ and positive for \numubar\ events. 

To reject NC interactions  we use a discriminant obtained by combining four event characteristics into a $k$NN variable~\cite{ref:RustemThesis}.  The selection criteria optimize selection efficiency and sample purity to obtain maximum sensitivity to oscillations~\cite{ref:ZeynepThesis,ref:ChrisThesis}.  Far Detector selection efficiencies for neutrino and antineutrino samples are 93\% and 97\%, with purities of 99\% and 94\%, respectively. Because the neutrino sample has a larger component of highly inelastic events, this optimization process reduces the optimal neutrino selection efficiency in favor of lower NC background. The lower overall purity of the antineutrino sample results from the much larger neutrino contamination at higher energies in the antineutrino mode; however, in the region of interest to oscillations and NSI the contamination is smaller. 

The FD neutrino and antineutrino spectra in the absence of flavor change are predicted using the ND data by first correcting the ND spectra for inefficiency and backgrounds and then extrapolating to the FD by a transfer matrix obtained from simulation~\cite{ref:minos2006,ref:JustinThesis}. We predict 2073 neutrino and 273 antineutrino events without oscillations, and 
observe 1654 and 193 events, respectively. 

\begin{figure}
\centering
\includegraphics[trim = 10mm 10mm 10mm 10mm, clip, width=\columnwidth]{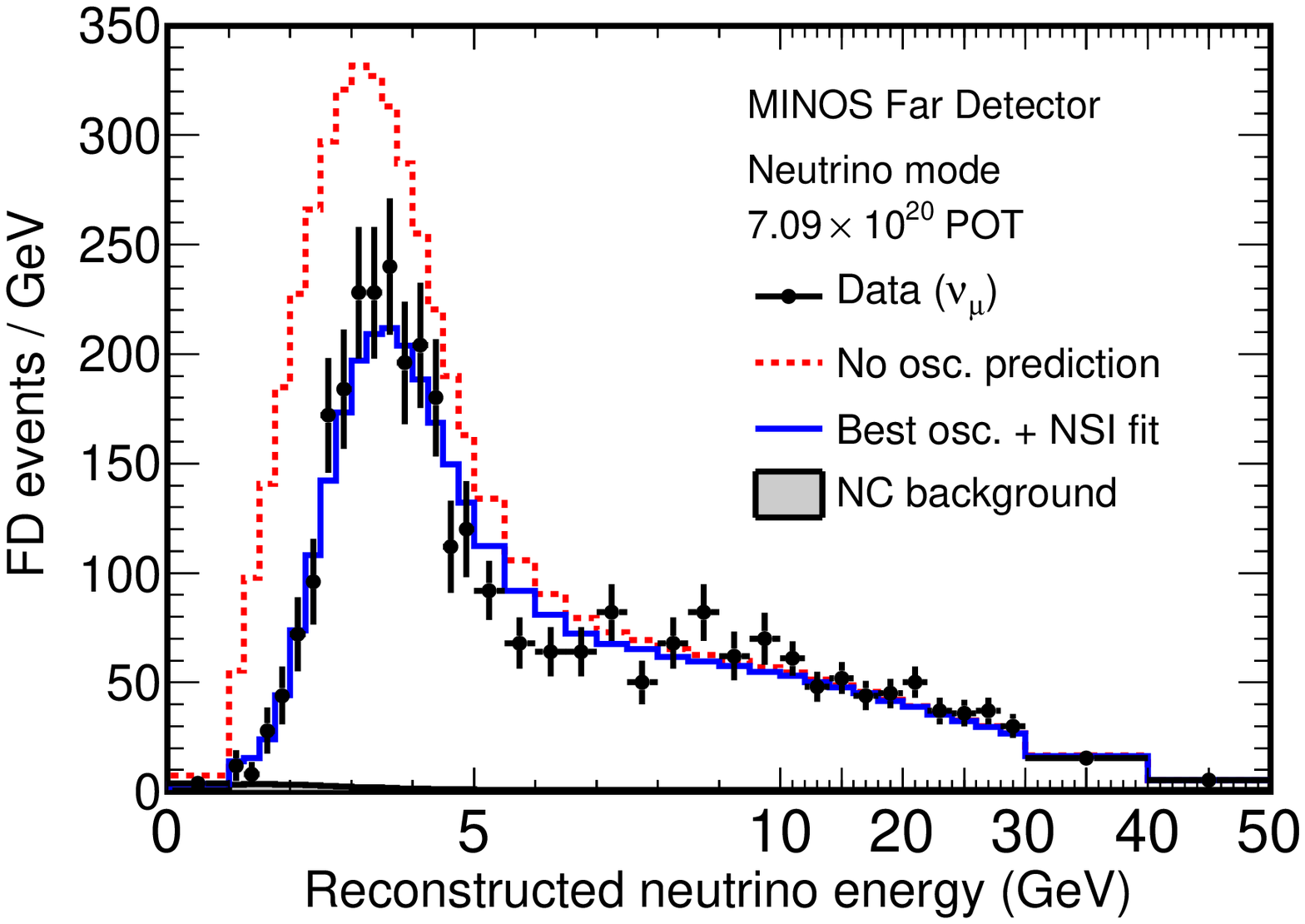}\\
\includegraphics[trim = 10mm 10mm 10mm 5mm, clip, width=\columnwidth]{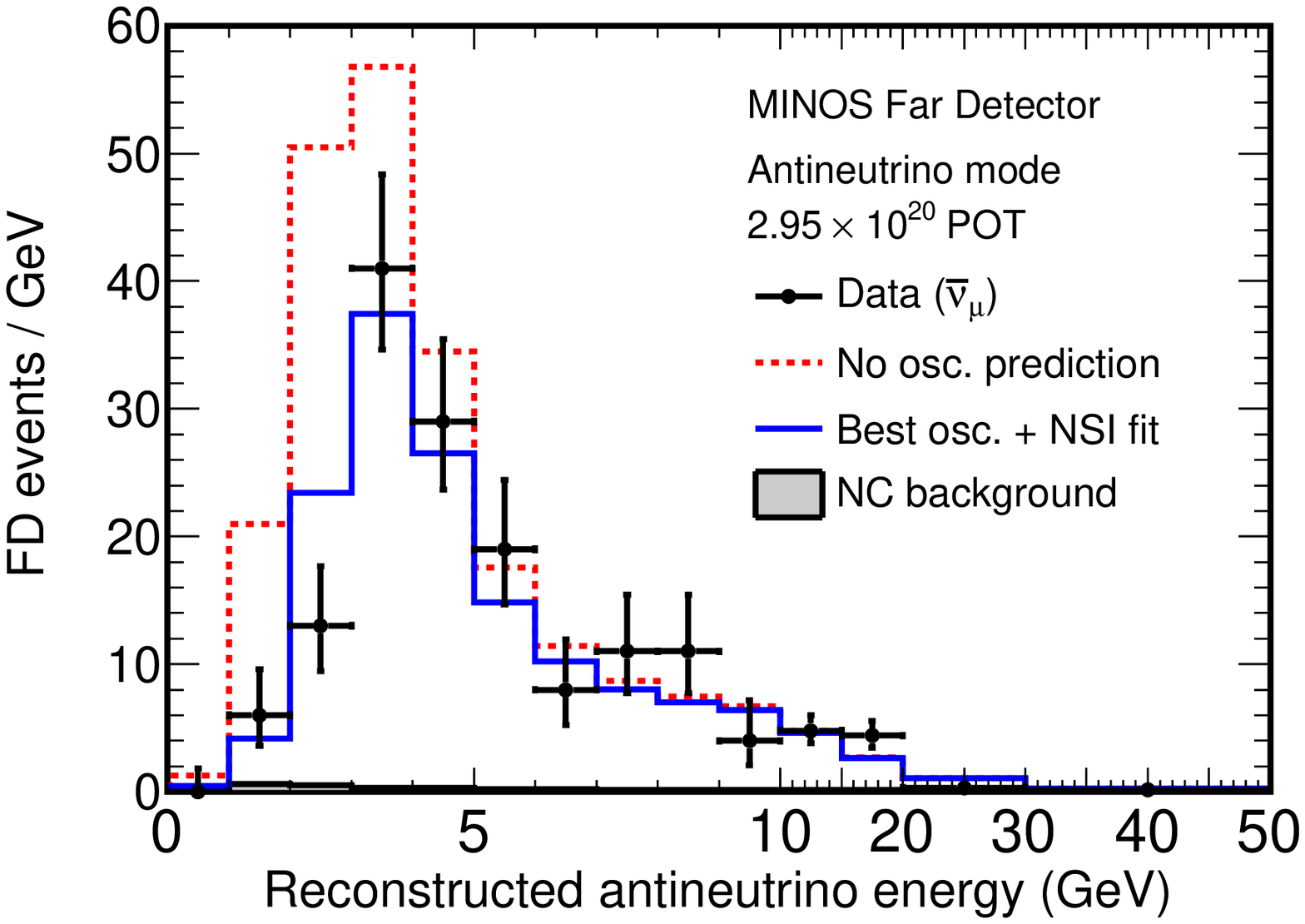}
\caption{Far Detector distributions of selected neutrino (top) and antineutrino (bottom) events. Black dots represent 
data, the dashed histogram shows the prediction in the absence of oscillations and the solid histogram shows 
the prediction for the model in Eq.~(\ref{eq:nsi_prob}) at the values obtained in our fit to the oscillation and NSI model. \label{fig:spectra}}
\end{figure}

The neutrino and antineutrino energy spectra are fit simultaneously to three parameters, \dmsq, \st, and \eps, in the 
combined oscillation and NSI model in Eq.~(\ref{eq:nsi_prob}), using a binned log-likelihood. The value of the mixing angle is constrained to be physical by asserting $0$~$\leq$~$\sin^{2}(2\theta)$~$\leq1$.  The resulting simulated energy spectra, obtained by using the best fit values, are shown in Fig.~\ref{fig:spectra} superimposed on the full neutrino and antineutrino spectra.

The overall systematic uncertainty in the measurement is much smaller than the statistical uncertainty. However, the difference in the relative sizes of the neutrino and antineutrino event samples results in a large difference in their statistical uncertainties, while the systematic uncertainties are comparable. Systematic uncertainties are included in the fit to ensure that the neutrino sample does not have a disproportionate impact. The largest sources of systematic uncertainty are the muon energy scale, the hadronic energy scale, the NC background, and the relative normalization between the Near and Far detectors. The muon energy scale uncertainty is 2\% for events where muon energy is determined from range and  3\% where the energy is obtained by measuring the curvature in the magnetic field. 

The uncertainty on the measurement from range is based on a comparison of tabulated muon spectrum power in MINOS detectors with tables in~\cite{ref:range}, and includes our uncertainty in the material composition of the detectors. The curvature uncertainty was found by comparing the curvature measurement to that from range for stopping muons~\cite{ref:minosCC2010}. The hadronic energy scale uncertainty is energy-dependent and smaller than 10\% integrated over all hadronic energies. Its multiple components arise from uncertainties in hadron production modeling and  from the uncertainty in detector response~\cite{ref:minosCC2010}. The NC background is less than 2\% of both the neutrino and antineutrino samples integrated across all energies. The size of its uncertainty, which is dominated by hadronic shower modeling, is estimated to be 20\% by comparing experimental data to simulation. The 1.6\% normalization uncertainty at all energies arises from modeling reconstruction differences between Near and Far detectors, and the uncertainty in the number of protons on target~\cite{ref:minosCC2010}.

These four sources of systematic uncertainty are included in the fit using penalty terms.
The best fit parameters from this procedure are found to be
\begin{gather}
  \Delta m^{2} = 2.39^{+0.14}_{-0.11}\times10^{-3}\text{eV}^{2}, \notag \\
  \sin^{2}(2\theta) = 1.00^{+0.00}_{-0.06}, \notag\\
  \varepsilon_{\mu\tau} = -0.07^{+0.08}_{-0.08},\notag
\end{gather}
\noindent with the allowed region $-0.20 < \varepsilon_{\mu\tau} < 0.07$ (90\% C.L.).

\begin{figure}[t]
  \begin{center}
    \includegraphics[trim = 10mm 10mm 10mm 10mm, clip, width=\columnwidth]{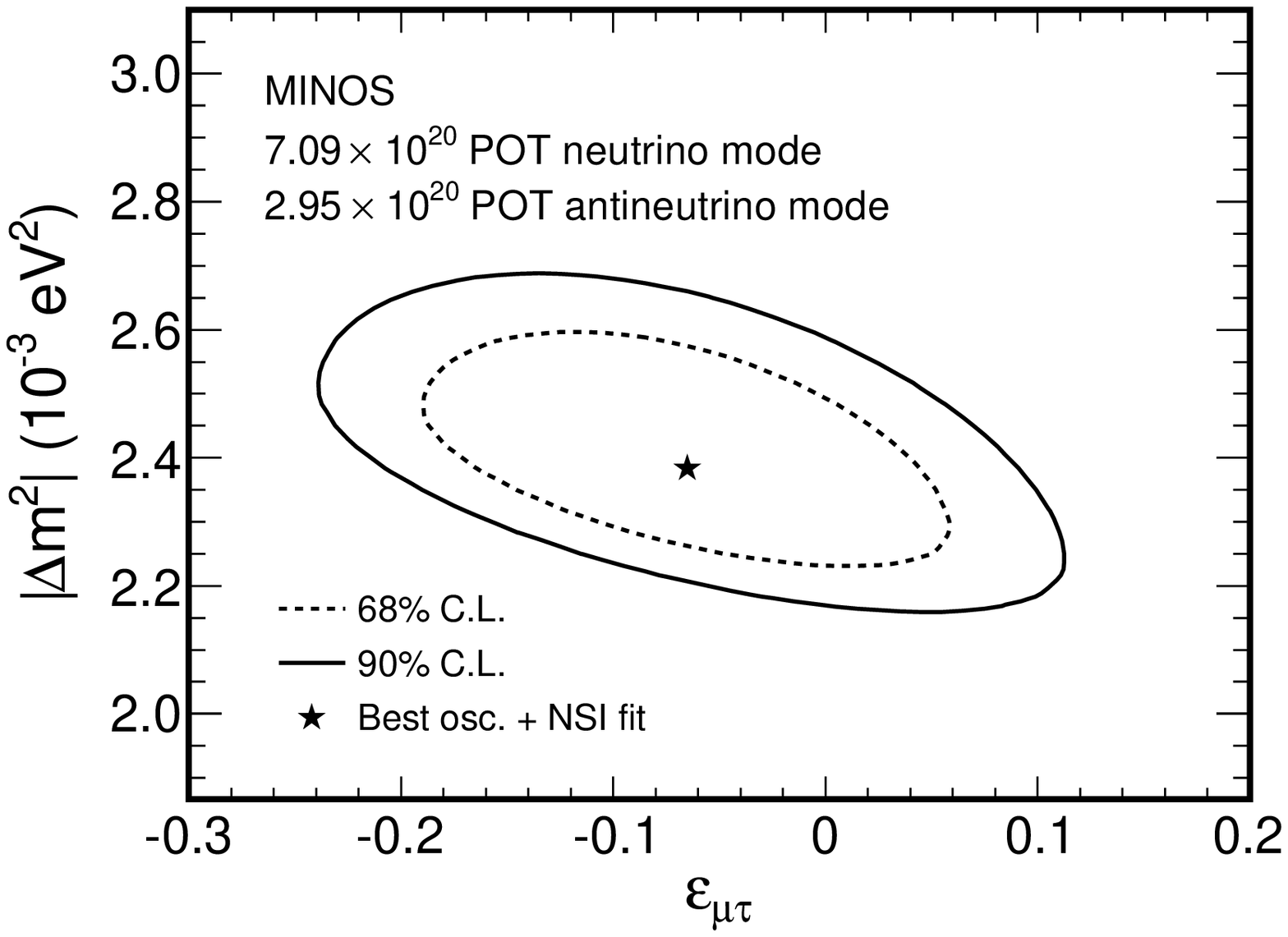}\\
    \includegraphics[trim = 10mm 10mm 10mm 10mm, clip, width=\columnwidth]{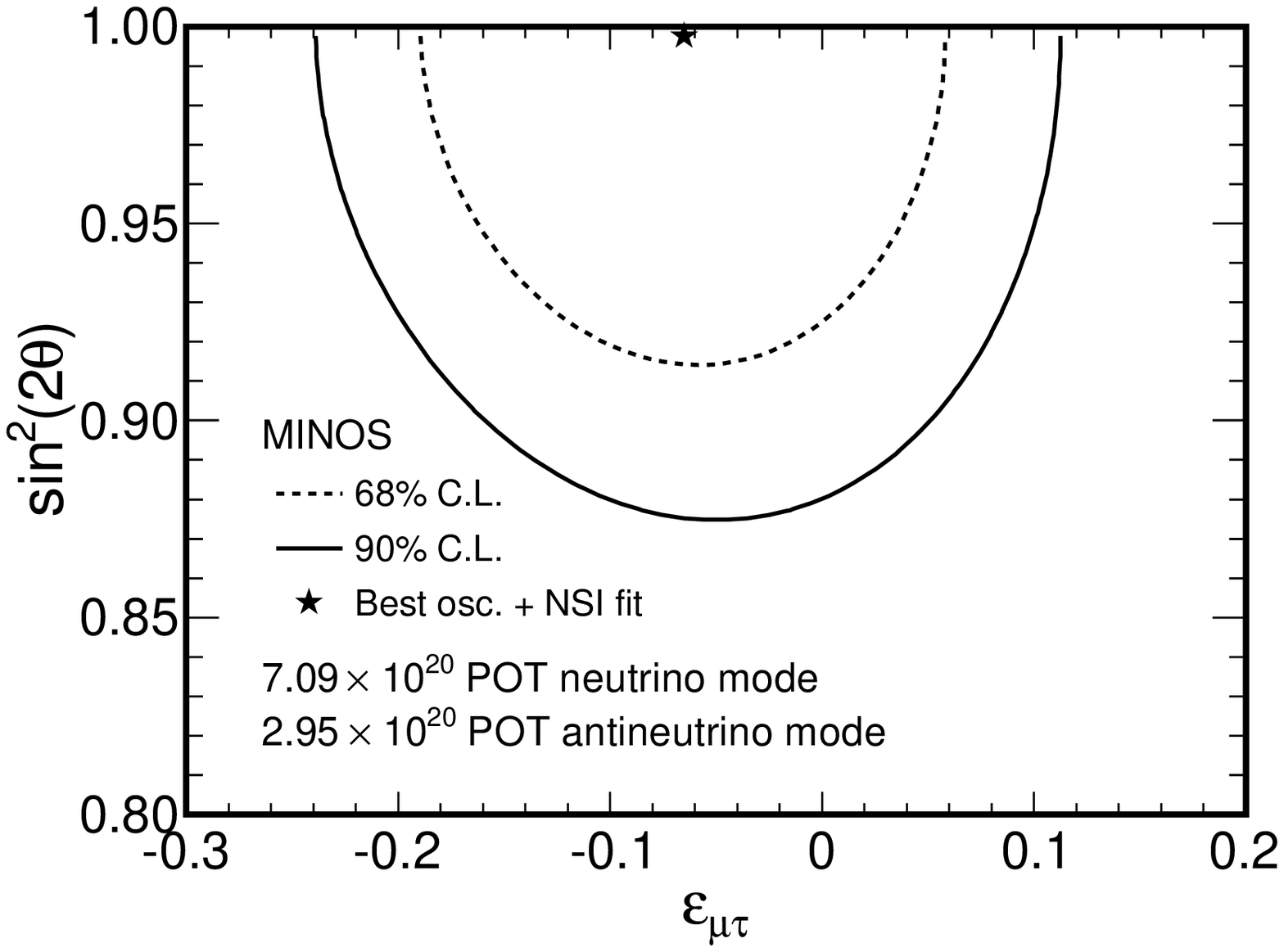}\\
    \includegraphics[trim = 10mm 10mm 10mm 10mm, clip, width=\columnwidth]{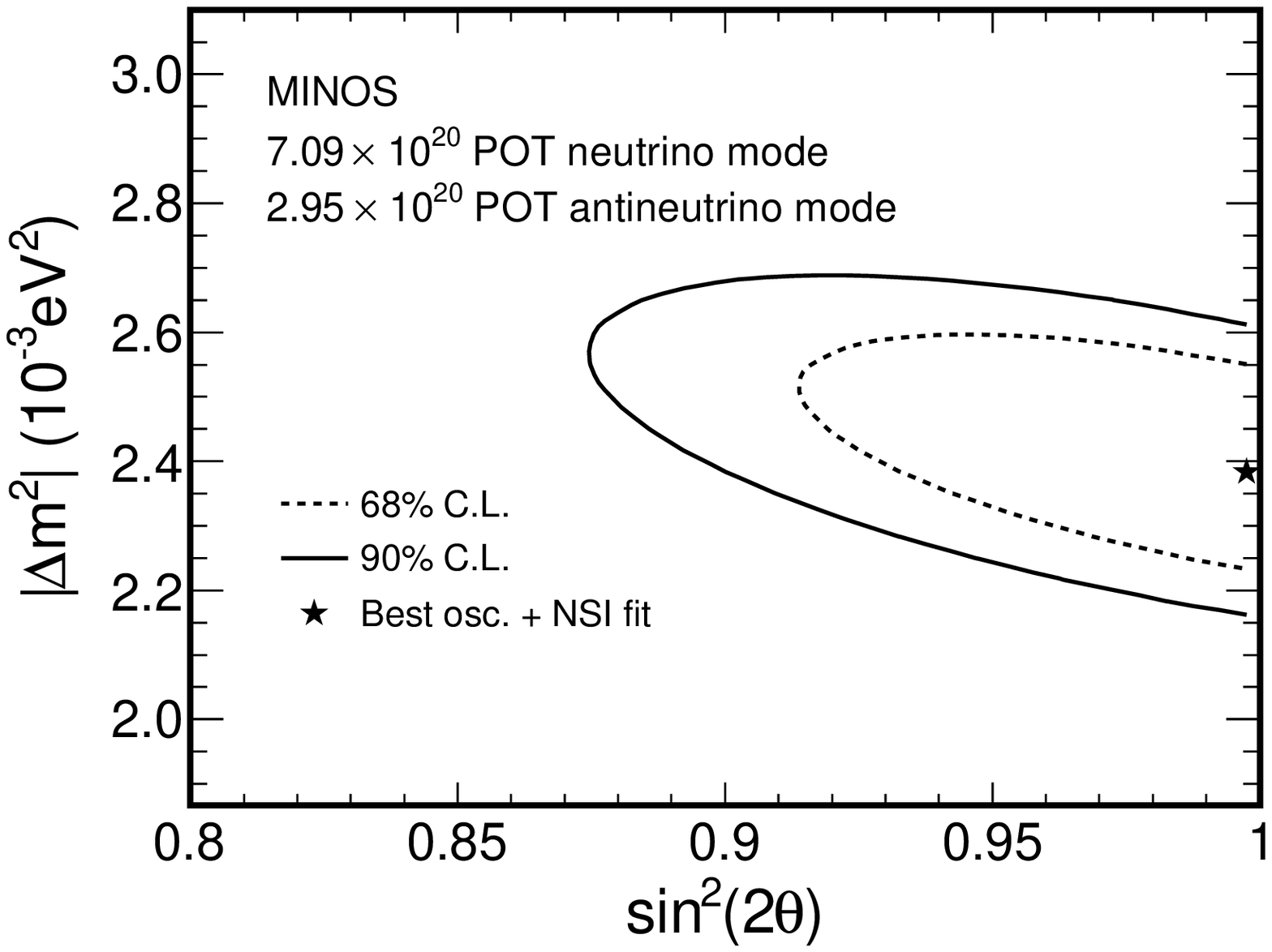}\\
    \caption{\label{fig:contours_nsi}The 68\% and 90\% allowed regions for the three parameters in the combined oscillation and NSI flavor change model given in Eq.~(\ref{eq:nsi_prob}). To obtain slices for each parameter pair combination we marginalize over the third parameter as well as over the systematic penalty parameters.}
  \end{center}
\end{figure}

The penalty terms from systematic uncertainties have a negligible effect on the fit; each penalty term pulls the best fit point by much less than one standard deviation. The  allowed regions of fit parameters are shown in Fig.~\ref{fig:contours_nsi}, where three two-dimensional slices from a 3D likelihood surface are chosen by marginalizing over the third parameter.  The obtained oscillation parameter values are in good agreement with previously published results in~\cite{ref:minosCC2010,ref:rhc}. Within errors the fit is consistent with no contribution to flavor change from NSI. This conclusion is in agreement with recent results from the Super-Kamiokande collaboration~\cite{ref:superK}, as well as with values of \eps\ extracted from global fits to data from multiple experiments~\cite{ref:blennow1,ref:blennow2,ref:biggio,ref:escrihuela}.

In summary, this is the first direct search for non-standard interactions with high-purity samples of both neutrinos and antineutrinos. We conducted a simultaneous fit to neutrino and antineutrino energy spectra of conventional $\nu_{\mu}\rightarrow\nu_{\tau}$ oscillations with an additional NSI matter effect. We found no evidence for non-standard neutrino interactions.

This work was supported by the U.S. DOE; the United Kingdom STFC; the U.S. NSF; the State and University of Minnesota; the University of Athens, Greece; and Brazil's FAPESP, CNPq and CAPES.  We are grateful to the Minnesota Department of Natural Resources, the crew of the Soudan Underground Lab, and the personnel of Fermilab for their contribution to this effort.


\begin{thebibliography}{99}
\bibitem{comment:dmbar} The experiment measures an unresolved mixture
  of $|\Delta m^{2}_{31}|$ and $|\Delta m^{2}_{32}|$, which is
  referred to as $|\Delta m^{2}|$ for brevity. The parameter
  $\sin^{2}(2\theta)$ is likewise an admixture, dominated by
  $\theta_{23}$. 
\end{thebibliography}
\end{document}